\documentclass{ws-ijmpd}
\usepackage[super,compress]{cite}
\usepackage{hyperref}
\hypersetup{colorlinks,urlcolor=black,citecolor=red,linkcolor=blue,filecolor=black}
\usepackage{breakurl}

\DeclareMathAlphabet{\mathpzc}{OT1}{pzc}{m}{it}
 
\def\beq{\begin{equation}}
\def\eeq{\end{equation}}
\def\bea{\begin{eqnarray}}
\def\eea{\end{eqnarray}}
\def\bwt{\begin{widetext}}
\def\ewt{\end{widetext}}
\def\nin{\noindent} 
\def\nn{\nonumber\\}

\begin{document}

\markboth{D.~G.~A.~Duniya}
{Qualitative probe of IDE with RSDs}

%
\catchline{}{}{}{}{}
%
\title{Qualitative probe of interacting dark energy\\ with redshift-space distortions}

\author{Didam~G.~A.~Duniya}
\address{Department of Physics \& Astronomy,\\ Botswana International University of Science and Technology,\\ Palapye, Botswana\\
duniyaa@biust.ac.bw}


\maketitle

\begin{history}
\received{24 January 2023}
\revised{23 October 2023}
\accepted{21 February 2024}
\end{history}

\begin{abstract}\nin
The imprint of interacting dark energy (IDE) needs to be correctly identified in order to avoid bias in constraints on IDE. This paper investigates the large-scale imprint of IDE in redshift space distortions, using \emph{Euclid}-like photometric prescriptions. A first attempt at incorporating the IDE dynamics in the galaxy (clustering and evolution) biases is made. Without IDE dynamics taken into account in the galaxy biases, as is conventionally done, the results suggest that for a constant dark energy equation of state parameter, an IDE model where the dark energy transfer rate is proportional to the dark energy density exhibits an alternating, positive-negative effect in the redshift space distortions angular power spectrum. However, when the IDE dynamics is incorporated in the galaxy biases, it is found that the apparent positive-negative alternating effect vanishes: implying that neglecting IDE dynamics in the galaxy biases can result in ``artefacts'' that can lead to incorrect identification of the IDE imprint. In general, the results show that multi-tracer analysis will be needed to beat down cosmic variance in order for the redshift space distortions angular power spectrum as a statistic to be a viable diagnostic of IDE. Moreover, it is found that redshift space distortions hold the potential to constrain IDE on large scales, at redshifts $z \,{\leq}\, 1$; with the scenario having IDE dynamics incorporated in the biases showing better potential.
\end{abstract}

\keywords{Interacting dark energy; redshift-space distortions; galaxy biases; angular power
spectrum.}



\section{Introduction}\label{sec:intro} 
An observational evidence \cite{Ferreira:2014jhn} for the existence of interacting dark energy (IDE) \cite{Ferreira:2014jhn,Kumar:2019wfs,DiValentino:2019jae,Yang:2014gza,Lucca:2021dxo,Costa:2016tpb,LopezHonorez:2010esq,Bhattacharyya:2018fwb,Costa:2019uvk,Chibana:2019jrf,Abdalla:2022yfr,Poulin:2022sgp,Borges:2023xwx,Duniya:2015nva,Duniya:2016gcf,Amendola:2003eq,Amendola:2003wa,Baldi:2011wy,Pace:2014tya,Maccio:2003yk,Xia:2009zzb,Clemson:2011an,Valiviita:2008iv,Hashim:2014rda,Duniya:2022xcz,Li:2018ydj,DiValentino:2019ffd,Halder:2021jiv,Kumar:2021eev,Wang:2021kxc,Nunes:2022bhn,Salvatelli:2013wra,Costa:2013sva} has been presented recently. This evidence was arrived at by analysing the data of the measurements of baryon acoustic oscillations (BAO) in the Ly$\alpha$ forest of the Baryon Oscillation Spectroscopic Survey (BOSS) DR11 quasars \cite{BOSS:2014hwf}. The BOSS results presented the first ever measurements of the evolution of dark energy (DE) at high redshifts ($z \,{=}\, 2.34$). The results indicated a $2.5\sigma$ deviation from the concordance model ($\Lambda$CDM): a universe dominated by only a cosmological constant $\Lambda$ and cold dark matter (CDM). The given deviation in the BAO measurements does not seem to be easily (or not at all) explained by common generalizations of the cosmological constant---which entails the standard, non-interacting dynamical DE. However, an alternative is IDE. Thus, by performing a global fitting of the cosmological parameters, with the combined data from BOSS and Planck \cite{Planck:2018vyg}, it was shown (see Ref.~\refcite{Ferreira:2014jhn}) that certain IDE models are able to predict results that are compatible with the ones obtained by BOSS.

This provides a strong step forward towards correctly identifying the imprint of IDE. Several other analyses have used various observations to test IDE models, in most of which cases the IDE models turn out to be compatible with the observations. Despite all these efforts, there is still no definitive answer from observational analysis or fundamental theory as to the true form of IDE. Although a lot of work has already been done, more needs to be done. The observational analyses need to go beyond background observables, to the perturbations. Moreover, current literature show that an in-depth qualitative analysis of the redshift space distortions (RSDs) with respect to IDE is still necessary, as previous works (e.g.~Refs.~\refcite{Ferreira:2014jhn}\nocite{Kumar:2019wfs,DiValentino:2019jae,Yang:2014gza,Lucca:2021dxo,Costa:2016tpb,LopezHonorez:2010esq,Bhattacharyya:2018fwb,Costa:2019uvk,Chibana:2019jrf,Abdalla:2022yfr,Poulin:2022sgp}--\refcite{Borges:2023xwx}) focus on the estimation of cosmological parameters.

In this paper, the imprint of IDE is probed with RSDs, on ultra-large scales, i.e.~scales near and beyond the Hubble radius. As at the time of compiling this paper, there does not appear to be any previous work that gives a detailed qualitative analysis of RSDs in IDE. Of the previous works cited above, none of the given analysis was done with respect to the RSDs angular power spectrum. (For works on constraints from RSDs, with respect to the angular power spectrum for non-IDE scenario, see e.g.~Ref.~\refcite{Fonseca:2019qek}.) Moreover, for the first time, the IDE dynamics is incorporated explicitly in the galaxy (clustering and evolution) biases. As galaxies cluster and evolve in a universe with IDE, the redshift-dependent galaxy biases should naturally acquire (directly) the dark-sector interaction. This has never previously been investigated. Thus, this paper focuses on providing an extended qualitative analysis of the imprint of IDE on ultra-large scales, aiming to highlight crucial insights on the nature of IDE: using the angular power spectrum.

In what follows, the dark-sector settings and models are outlined in Sec.~\ref{sec:IDE}. In Sec.~\ref{sec:Delta-RSD}, the RSDs (overdensity and angular power spectrum) are outlined and investigated: with, and without, the IDE dynamics incorporated in the galaxy biases. Lastly, a conclusion is given in Sec.~\ref{sec:Concl}.

\vfill


\section{The Dark Universe}\label{sec:IDE}

Henceforth, the late-time Universe with anisotropic-stress-free spacetime metric, is assumed; the metric:
\bea\label{metric}
ds^2 = a^2 \left[-(1+2\Phi) d{\eta}^2 + (1-2\Phi) d{\bf x}^2 \right],
\eea
where $a \,{=}\, a(\eta)$ is the scale factor, $\eta$ is the conformal time, $\Phi \,{=}\, \Phi({\bf x},\eta)$ is the Bardeen potential \cite{Bardeen:1980kt}, and ${\bf x}$ is the spatial coordinate. 

\subsection{The IDE description}
In a universe dominated by (cold) dark matter (DM) and DE, the background evolution equations for the cosmic species $A$, are given by
\bea\label{conservatn}
\bar{\rho}_A' + 3{\cal H}(1+w_{A,{\rm eff}}) \bar{\rho}_A = 0,
\eea
where $A \,{=}\, m$ for DM, and $A \,{=}\, x$ for DE; $\bar{\rho}_A$ is the energy density for $A$, an overbar denotes background, a prime denotes derivative with respect to $\eta$, and ${\cal H} = a'/a$ is the conformal Hubble parameter, and
\beq\label{eff:EoS}
w_{A,{\rm eff}} \;\equiv\; w_A -\dfrac{a\bar{Q}^{(I)}_A}{3{\cal H}\bar{\rho}_A},
\eeq
with $w_m \,{=}\, \bar p_m/\bar\rho_m \,{=}\, 0$ and $w_x \,{=}\, \bar{p}_x / \bar{\rho}_x$ being the equation of state parameters; $\bar{p}_A$ is the pressure for $A$. Moreover, the conservation of the total energy-momentum tensor implies that the background energy (density) transfer rates are given by
\beq\label{QxQm}
\bar{Q}^{(I)}_m = {-}\bar{Q}^{(I)}_x,
\eeq 
where the superscript $I$ denotes the model type (see Sec.~\ref{subsec:Q}); with $\bar{Q}^{(I)}_x \,{>}\, 0$ corresponding to energy transfer from DM to DE (and vice versa). 

Here DE is taken as a fluid with constant $w_x$; with the transfer $4$-vectors $Q^\mu_A$ being parallel to the DE $4$-velocity:
\beq\label{trans:Case}
Q^\mu_x = Q^{(I)}_x u^\mu_x = -Q^\mu_m ,
\eeq
where this implies that there is zero momentum transfer in the DE rest frame (see e.g.~Refs.~\refcite{Duniya:2015nva}--\refcite{Xia:2009zzb}), and we have
\beq\label{overDens:Vels}
u^\mu_A = a^{-1}\left(1 -\Phi,\, \partial^i V_A\right),\; u^\mu = a^{-1}\left(1 -\Phi,\, \partial^i V\right),
\eeq
where $\Phi$ is as given by \eqref{metric}, $V_A$ is the velocity potential for $A$; with $V$ being the total velocity potential. Thus, we have the momentum density transfer rates:
\bea\label{fm:fx}
f^{(I)}_x = \bar{Q}^{(I)}_x (V_x - V) = -f^{(I)}_m.
\eea

\subsection{The IDE models}
\label{subsec:Q}
Two phenomenological models for the DE energy transfer rate are adopted in this work. The goal is not to compare the models, but to probe the nature of IDE in two widely-investigated models in the literature. 

The first model is given by \cite{Duniya:2015nva,Duniya:2016gcf,Clemson:2011an,Valiviita:2008iv,Hashim:2014rda},
\beq\label{Mod1:Q}
Q^{(1)}_x \equiv \Gamma \rho_x = \bar{Q}^{(1)}_x (1 +\delta_x),
\eeq
where $\bar{Q}^{(1)}_x \,{=}\, \Gamma\bar{\rho}_x \,{=}\, {-}\bar{Q}^{(1)}_m$ is the background term, with the second equality being prescribed by \eqref{QxQm}; $\Gamma$ is the interaction strength (a constant), with $\Gamma \,{<}\, 0$ corresponding to a decay of DE into DM, and $\delta_x \,{\equiv}\, \delta\rho_x/\bar{\rho}_x$ is the DE density contrast.

The second model is \cite{Duniya:2015nva, Duniya:2016gcf, Duniya:2022xcz}
\begin{align}\label{Mod2:Q}
Q^{(2)}_x \equiv &\; \dfrac{1}{3} \xi \rho_x \nabla_\mu u^\mu ,\nn
=&\; \bar{Q}^{(2)}_x \Big[1 +\delta_x -\Phi -\dfrac{1}{3{\cal H} }\left(3\Phi' - \nabla^2 V\right)\Big],
\end{align}
where $\bar{Q}^{(2)}_x \,{=}\, \xi{\cal H}\bar{\rho}_x/a \,{=}\, {-}\bar{Q}^{(2)}_m$ is the background term, with the second equality being prescribed by \eqref{QxQm} and, $\xi$ is the interaction strength (a constant). 

It should be pointed out that $Q^{(2)}_x$ has never previously been investigated for redshift space distortions. However, there is a popular approximation of $Q^{(2)}_x$ in the literature (see e.g.~Refs.~\refcite{Kumar:2019wfs}--\refcite{Lucca:2021dxo},~\refcite{Li:2018ydj}--\refcite{Nunes:2022bhn}), given by $\hat{Q}_x \,{\equiv}\, a^{-1} \xi{\cal H} \rho_x \,{=}\, \bar{Q}^{(2)}_x \left(1 +\delta_x\right)$, where $\bar{Q}^{(2)}_x$ is as given by~\eqref{Mod2:Q} and, $\delta_x$ given by~\eqref{Mod1:Q}. We notice that, in the background, $\hat{Q}_x$ is identical to $Q^{(2)}_x$; hence, will give the same background cosmology. The main difference of the two IDE models is in the perturbations, which should lead to divergent evolution in the perturbations. The $Q^{(2)}_x$ model is driven by the total (background and perturbations) expansion rate, $\frac{1}{3}\nabla_\mu u^\mu$, whereas $\hat{Q}_x$ is limited to the background (Hubble) expansion rate, ${\cal H}$. Consequently, $Q^{(2)}_x$ and $\hat{Q}_x$ will give different large-scale cosmology; with the two IDEs giving different imprints in large scale structure. (A comparison of $\hat{Q}_x$ and $Q^{(2)}_x$ is left for future work.) 

The range of $w_x$ is restricted by the stability requirements, given by \cite{Duniya:2015nva, Duniya:2016gcf, Clemson:2011an, Salvatelli:2013wra, Costa:2013sva}
\beq\label{stab}
w_x > -1~~\mbox{for}~\xi,\Gamma > 0; ~~w_x < -1~~\mbox{for}~\xi,\Gamma < 0,
\eeq
for a constant $w_x$. These two cases correspond to different energy transfer directions, by \eqref{Mod1:Q} and \eqref{Mod2:Q}:
\beq\label{etd}
\mbox{DM $\to$ DE for}~\xi,\Gamma > 0; ~\mbox{DE $\to$ DM for}~\xi,\Gamma < 0.
\eeq
In this work, \mbox{DM $\to$ DE} transfer direction is adopted. This ensures that DE has more accelerating power, and can admit $w_{x\rm eff} \,{<}\, {-}1$, i.e. the IDE behaves like an uncoupled ``phantom'' DE, but without the problems \cite{Huey:2004qv} associated with phantom DE \cite{Valiviita:2008iv}. Thus, this choice can conveniently accommodate the cosmology of (uncoupled) phantom DE models. (See e.g.~Refs.~\refcite{Duniya:2015nva,Clemson:2011an,Valiviita:2008iv,Salvatelli:2013wra} and~\refcite{Costa:2013sva} for the evolution equations.)


\section{Redshift Space Distortions}\label{sec:Delta-RSD}

\subsection{The overdensity}
\label{subsec:Dens}
The RSDs act to squash overdense regions and amplify underdense regions, eventually boosting the galaxy overdensity: this effect is corrected for by the shear field or gradient of the peculiar velocity field, given by \cite{Kaiser:1987qv, Strauss:1996eq, Hamilton:1997zq, Assassi:2017lea, Bonvin:2014owa}
\bea\label{Delta-RSD}
\Delta_n({\bf n},z) \;=\; \Delta_{\rm g}({\bf n},z) - \dfrac{1}{{\cal H}} \dfrac{\partial}{\partial \bar{r}} V_\parallel({\bf n},z), 
\eea
where the signal is observed to propagate in the direction ${-}{\bf n}$ at (background) redshift $z$, and $\Delta_{\rm g}$ is the galaxy comoving overdensity; with the second term in \eqref{Delta-RSD} correcting for RSDs, $\bar{r} \,{=}\, \bar{r}(z)$ being the background radial comoving distance at $z$ and, $V_\parallel \,{=}\, {-}{\bf n}\,{\cdot} {\bf V}_{\rm g}$ being the line-of-sight peculiar velocity component; ${\cal H}$ is as in Sec.~\ref{sec:IDE}.

In reality, RSDs do exist in the large-scale structure among several effects other than the matter density; these include the Doppler and the potential effects (neglecting integral effects, which are negligible at $ z \,{\lesssim}\, 1$, being the $z$ of interest in this work). These effects together (apart from density and RSDs) are, henceforth, termed ``local effects.'' They surface in the overdensity in redshift space through the redshift and the volume perturbations, accordingly. The local effects typically amount to a relatively small contribution compared to the terms in \eqref{Delta-RSD}. Moreover, \eqref{Delta-RSD} is sufficient to obtain a true understanding of the nature of RSDs in the angular power spectrum. However, these effects need to be taken into account in the light of upcoming precision cosmological era.

Thus, from \eqref{Delta-RSD}, the corrected overdensity becomes 
\beq\label{Delta-full}
\Delta_n({\bf n},z) \;\to\; \Delta_n({\bf n},z) + \Delta_{\rm loc}({\bf n},z), 
\eeq
where the local term, is given by
\begin{align}\label{Delta-loc}
\Delta_{\rm loc}({\bf n},z) \equiv & \left(b_e - \dfrac{{\cal H}'}{{\cal H}^2} - \dfrac{2}{{\cal H}\bar{r}}\right) V_\parallel({\bf n},z) + \dfrac{1}{{\cal H}}\Phi'({\bf n},z) \nn
&+\; \left(\dfrac{{\cal H}'}{{\cal H}^2} + \dfrac{2}{{\cal H}\bar{r}} - 1 - b_e\right) \Phi({\bf n},z) + \left(3 - b_e\right){\cal H}V({\bf n},z), 
\end{align}
where ${\cal H}$, $\bar{r}$, $V$, $V_\parallel$, and $\Phi$ are as previously given; with $b_e \,{=}\, b_e(z)$ being the (galaxy) evolution bias, given by
\beq\label{b_e}
b_e \equiv \dfrac{\partial \ln(a^3\bar{n}_{\rm g})}{\partial\ln(a)} ,
\eeq
where $\bar{n}_{\rm g}$ is the galaxy number per unit volume.

Observers are able to measure the background number of sources $\bar{N}$ per unit solid angle $\Omega$ per redshift:
\beq\label{dNdzdO}
\dfrac{d\bar{N}(z)}{dzd\Omega} = \bar{n}_{\rm g}(z) \bar{\cal V}(z) \equiv \bar{\cal N}(z),
\eeq
where $\bar{\cal V} \,{=}\, \partial\bar{\upsilon}/ (\partial{z}\partial\Omega)$ is the volume $\bar{\upsilon}$ per unit solid angle per redshift. By using \eqref{dNdzdO}, the evolution bias \eqref{b_e} becomes
\beq\label{b_e2}
b_e = \dfrac{2a^2}{{\cal H}\bar{r}} + \dfrac{\partial\ln\bar{\cal N}}{\partial\ln(a)} + \dfrac{{\cal H}'}{{\cal H}^2} - 1 ,
\eeq
where $\bar{\cal V} \,{=}\, a^4\bar{r}^2 {\cal H}^{-1}\sin\theta_O$ (see e.g.~Ref.~\refcite{Duniya:2016ibg} for details); with $\theta_O$ being the angle at the observer.

\subsection{The angular power spectrum}\label{subsec:Cls}

In practice, observers often split galaxy surveys into $z$ bins, and the angular power spectra are measured either within the same bin, or as cross-correlations between two different bins centred at $z \,{=}\, z_S$ and $z \,{=}\, \hat{z}_S$: 
\begin{align}\label{Cls} 
C_\ell(z_S,\hat{z}_S) = \dfrac{4}{\pi^2} \left(\dfrac{9}{10}\right)^2 \int dk\, k^2 T(k)^2 P_{\Phi_p}(k) \Big| F_\ell(k,z_S) F^*_\ell(k,\hat{z}_S) \Big| ,
\end{align}
where $T(k)$ is the linear transfer function (linking the primordial fluctuations to the late-time perturbations), $P_{\Phi_p}$ is the gravitational potential field spectrum, and
\beq\label{F_ell}
F_\ell(k,z_S) \equiv \int^\infty_0 dz\, W(z_S,z) f_\ell(k,z),
\eeq
with $W$ being a window function---which gives the probability distribution of the sources in a given $z$ bin---here taken as a Gaussian, given by
\beq\label{windowfunc}
W(z_S,z) \;=\; \dfrac{1}{\sqrt{2\pi}\sigma_z} \exp{\left[-\dfrac{\left(z_S - z\right)^2}{2\sigma^2_z}\right]} ,
\eeq 
where $z_S$ is the mean redshift of the bin containing $z$, with $\sigma_z$ being the standard diviation from $z_S$; and 
\begin{align}\label{f_ell}
f_\ell(k,z) =\; & b\check{\Delta}_m(k,z) j_\ell(k\bar{r}) - \dfrac{1}{{\cal H}} \dfrac{\partial}{\partial\bar{r}} \check{V}^\parallel_m (k,z) \dfrac{\partial^2 j_\ell(k\bar{r})}{\partial(k\bar{r})^2} + \left(b_e - \dfrac{{\cal H}'}{{\cal H}^2} - \dfrac{2}{{\cal H}\bar{r}}\right) \nn
&\times \check{V}^\parallel_m(k,z) \dfrac{\partial j_\ell(kr)}{\partial(k\bar{r})} + \Big[(3 - b_e){\cal H}\check{V}_m(k,z) + \dfrac{1}{{\cal H}}\check{\Phi}'(k,z) \Big] j_\ell(k\bar{r}) \nn
&+ \left(\dfrac{{\cal H}'}{{\cal H}^2} + \dfrac{2}{{\cal H}\bar{r}} - 1 - b_e\right) \check{\Phi}(k,z) j_\ell(k\bar{r}) .
\end{align} 
where $b \,{=}\, b(z)$ is the clustering bias \cite{Desjacques:2016bnm, Duniya:2016ibg, Jeong:2011as} (well-known as ``galaxy'' bias; see~Ref.~\refcite{Desjacques:2016bnm} for an extensive review), $j_\ell$ is the spherical Bessel function and, $\check{\Delta}_m$ and $\check{V}_m^\parallel \,{=}\, \partial\check{V}_m/\partial{\bar{r}}$ are the matter comoving overdensity and the line-of-sight matter peculiar velocity component, respectively, divided by the gravitational potential $\Phi(k,z_d)$ (see e.g.~Refs.~\refcite{Duniya:2015nva,Duniya:2016ibg} and~\refcite{Duniya:2019mpr}) at the decoupling epoch $z \,{=}\, z_d$. The plane-parallel approximation \cite{Kaiser:1987qv}, which is suitable for the linear power spectrum, is avoided in \eqref{f_ell}. We used that on linear scales (being the scales in consideration here) galaxies follow the same path as the underlying matter: $V^\parallel_{\rm g} \,{=}\, V^\parallel_m$.

A value of the DE equation of state parameter, $w_x = {-}0.8$, is adopted for all numerical computations; also assuming Euclid photometric empirical fittings \cite{EuclidTheoryWorkingGroup:2012gxx}:
\begin{align}\label{b-fit}
b(z) =&\; \sqrt{1+z},\\ \label{N-fit}
\bar{\cal N}(z) =&\; 3.5\,{\times}\,10^8 z^2 \exp\left[-\left(\dfrac{z}{z_0}\right)^{3/2}\right] , 
\end{align}
where $z_0 \,{=}\, z_{\rm mean}/1.412$; with $z_{\rm mean} \,{=}\, 0.9$, and the evolution bias \eqref{b_e2} is computed using \eqref{N-fit}. 

\begin{figure*}\centering
\includegraphics[scale=0.3]{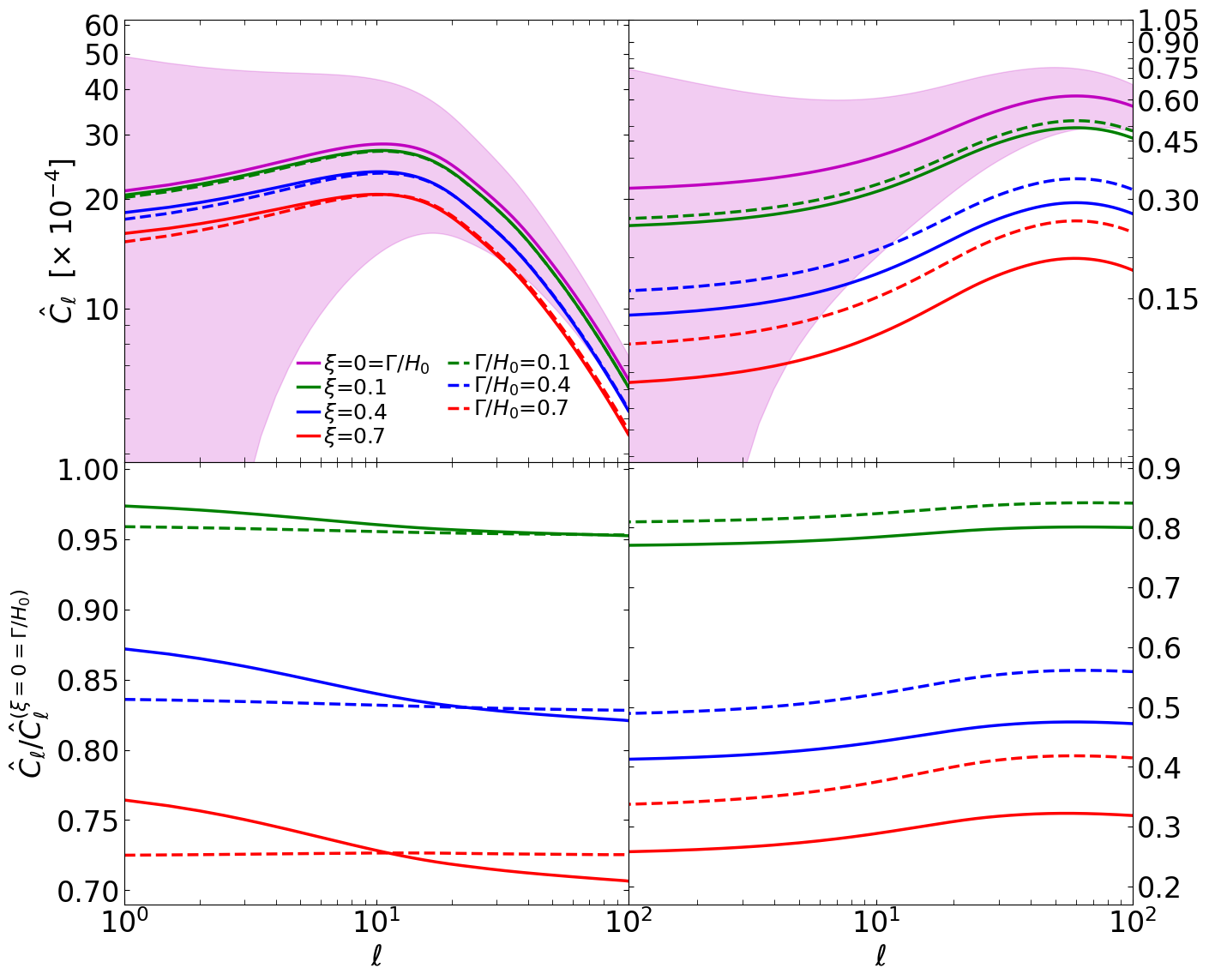}
\caption{\emph{Top:} The plots of the RSDs angular power spectrum $\hat{C}_\ell \,{=}\, C_\ell^{(\Delta_{\rm loc} {=}\, 0)}$ as a function of multipole $\ell$, at fixed source redshifts $\hat{z}_S \,{=}\, z_S \,{=}\, 0.1$ (left) and $\hat{z}_S \,{=}\, z_S \,{=}\, 1$ (right). The line styles denote IDE models $Q^{(1)}_x$ (dashed lines) and $Q^{(2)}_x$ (solid lines). The shaded regions denote cosmic variance \eqref{cosmicVar}. \emph{Bottom:} The plots of the corresponding ratios of $\hat{C}_\ell$, with IDE ($\Gamma/H_0 \,{\neq}\, 0$ and $\xi \,{\neq}\, 0$) to that with no IDE ($\Gamma/H_0 \,{=}\, 0 \,{=}\, \xi$), as a function of multipole $\ell$: at the same source redshifts. The line colours and styles are the same througout, and denote values of the IDE interaction strengths $\xi$ and $\Gamma/H_0$, as shown in the top-left legend.}\label{fig:clsWRTlTotalIDEeffect}
\end{figure*}

Background initial conditions for each IDE model were chosen for the cosmological equations to obtain the same matter density parameter $\Omega_{m0} \,{=}\, 0.24$ and Hubble constant $H_0 \,{=}\, 73~{\rm km\, s^{-1}\, Mpc^{-1}}$ as its corresponding standard (non-interacting) model. Thus, the IDE models are ``normalized'' to the same background universe at present epoch ($z \,{=}\, 0$). An advantage of this is that any scale-dependent deviations from standard behaviour will be isolated on ultra-large scales, at present epoch. Specifically, the power spectra for all the chosen values of $\xi$ and $\Gamma/H_0$ will merge on small scales at present epoch, and diverge otherwise. (At earlier epochs, $z \,{>}\, 0$, the power spectra will diverge on small scales.) Adiabatic initial conditions (see e.g.~Refs.~\refcite{Duniya:2015nva} and~\refcite{Duniya:2019mpr}) were adopted for the perturbations equations. The angular power spectrum \eqref{Cls} was computed by employing \eqref{b_e2} and \eqref{F_ell}--\eqref{N-fit}. 

\begin{figure*}\centering
\includegraphics[scale=0.32]{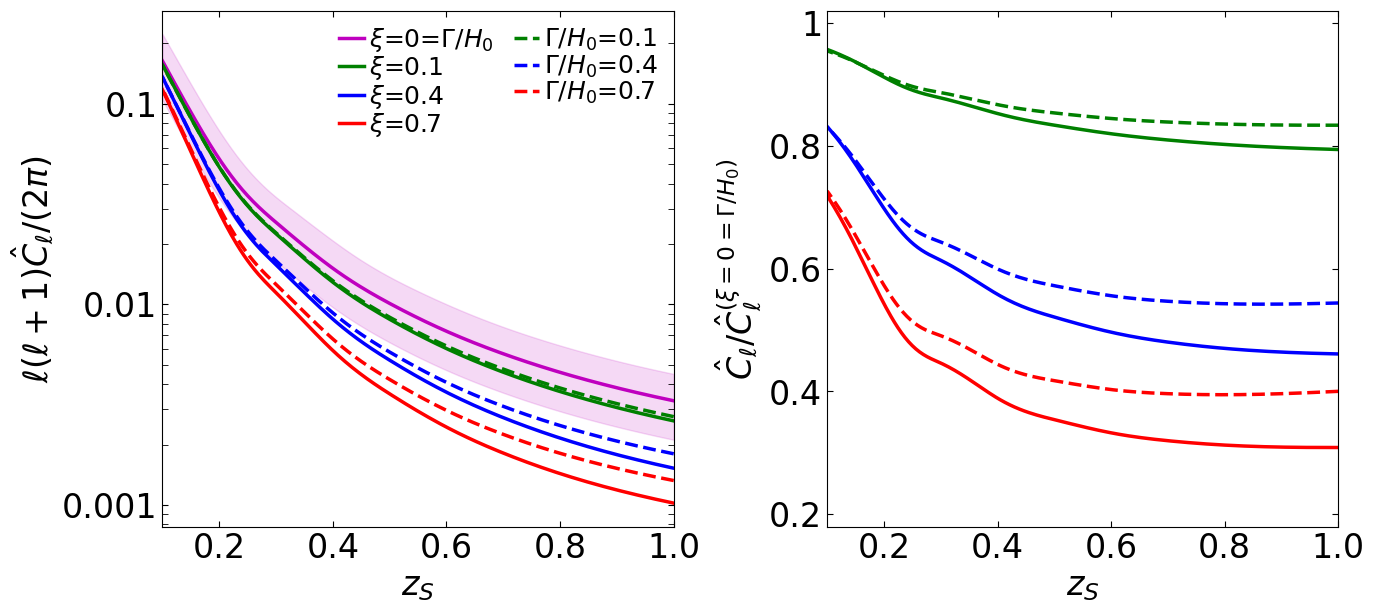}
\caption{\emph{Left:} The plots of the RSDs angular power spectrum $\hat{C}_\ell$ as a function of source redshift $z_S$ ($0.1 \leq z_S \leq 1$), at fixed multipole $\ell \,{=}\, 20$. \emph{Right:} The plots of the ratios of $\hat{C}_\ell(z_S)$ in the left panel, with IDE ($\Gamma/H_0 \,{\neq}\, 0$ and $\xi \,{\neq}\, 0$) to that with no IDE ($\Gamma/H_0 \,{=}\, 0 \,{=}\, \xi$), as a function of $z_S$, for the same value of $\ell$ and $z_S$ interval. All notations are as in Fig.~\ref{fig:clsWRTlTotalIDEeffect}.}\label{fig:clsWRTzTotalIDEeffect}
\end{figure*}

Fig.~\ref{fig:clsWRTlTotalIDEeffect} (top panels) shows plots of the RSDs angular power spectrum $\hat{C}_\ell \,{=}\, C^{(\Delta_{\rm loc} {=} 0)}_\ell$ as a function of multipole $\ell$, for $Q^{(1)}_x$ and $Q^{(2)}_x$: at the source redshift values $\hat{z} \,{=}\, z_S \,{=}\, 0.1$ and $\hat{z}_S \,{=}\, z_S \,{=}\, 1.0$. We see that at both redshifts, there is a consistent decrease in the angular power with increase in the IDE  interaction strengths $\Gamma/H_0$ and $\xi$, accordingly. This is not surprising since we chose the direction of energy-momentum transfer to be from DM to DE. Moreover, we see the relative qualitative effect of the two IDE models: $Q^{(1)}_x$ leads to larger supression, albeit marginal, in $\hat{C}_\ell$ (on the largest scales) at $z_S \,{=}\, 0.1$; whereas, at $z_S \,{=}\, 1$, $Q^{(2)}_x$ leads to more supression. We see that at $z_S \,{=}\, 0.1$, although both $Q^{(1)}_x$ and $Q^{(2)}_x$ lead to power suppression, there is more apparent separation between the models on larger scales than on smaller scales for each value of the interaction strengths ($\xi$ and $\Gamma/H_0$), and at $z_S \,{=}\, 1$, the two models diverge on all scales and increasingly so with increasing interaction strengths. This is a consequence of our normalization, where the IDE model are equalized on the background at present epoch. Hence small-scale behaviours will tend to converge at $z \,{\to}\, 0$; with the large-scale behaviours remaining only minimally (or not) affected.

It is well known that cosmic variance $\sigma^2_\ell$, given by
\bea\label{cosmicVar}
\sigma_\ell(z) = \sqrt{\dfrac{2}{\left(2\ell+1\right)f_{\rm sky}}}\, \hat{C}_\ell(z),
\eea
becomes substantial on the largest scales. (We adopt a value of the sky fraction,  $f_{\rm sky} \,{=}\, 0.375$, as covered by a Euclid photometric survey.) Thus, the extend of $\sigma_\ell$ is shown in Fig.~\ref{fig:clsWRTlTotalIDEeffect} (shaded regions, top panels), for (standard) non-IDE. The results imply that changes induced by IDE, described by $Q^{(1)}_x$ or $Q^{(2)}_x$, will ordinarily be overshadowed by cosmic variance and hence will not be observed by single tracers, e.g. galaxies with the same bias, except for high interaction strengths ($\xi, \Gamma/H_0 \gtrsim 0.4$) at higher source redshifts ($z_S \,{\gtrsim}\, 0.5$) and scales $\ell \,{\gtrsim}\, 20$. However, the values $(\xi, \Gamma/H_0) \gtrsim 0.4$ seem to be ouside observational constraints (see e.g.~Refs.~\refcite{Yang:2014gza,Clemson:2011an,Li:2018ydj,Kumar:2021eev} and~\refcite{Wang:2021kxc}); albeit not completely ruled out (see e.g.~Ref.~\refcite{Kumar:2019wfs}). Nevetheless, it should be noted that the given cited constraints are computed in the context of single tracers. However, multi-tracer techniques (see e.g.~Refs.~\refcite{Fonseca:2015laa}\nocite{Alonso:2015sfa,Witzemann:2018cdx}--\refcite{Qi:2021iic}) can be used to beat down $\sigma_\ell$ with future surveys, and hence provide the potential of detecting the IDE effects. 

Similarly, Fig.~\ref{fig:clsWRTzTotalIDEeffect} (left panel) shows plots of the RSDs angular power spectrum $\hat{C}_\ell$ as a function of source redshift $z_S$ (where $\hat{z}_S \,{=}\, z_S$), for a multipole $\ell \,{=}\, 20$. The extent of $\sigma_\ell$ is also indicated. Basically, Fig.~\ref{fig:clsWRTzTotalIDEeffect} (left panel) gives the evolutionary behaviour of the RSDs angular power spectrum. We see that the amplitude of $\hat{C}_\ell(z_S)$ falls steeply with increasing $z_S$. Moreover, similar to the results in Fig.~\ref{fig:clsWRTlTotalIDEeffect}, we see that the dark sector interactions lead to a suppression in the angular power, at the given range of $z_S$: this suppression follows from our choice of the direction of energy-momentum flow, being DM to DE. We also see the extent of the cosmic variance (shaded region), which is very slowly increasing with increasing source redshift. This is because cosmic variance is mainly a scale-dependent quantity, and only indirectly redshift-dependent via the angular power spectrum. The results also complement those in Fig.~\ref{fig:clsWRTlTotalIDEeffect} (top panels): for a single-tracer analysis the observable signal will ordinarily be overshadowed by cosmic variance for smaller interaction strengths ($\xi, \Gamma/H_0 \,{<}\, 0.4$) and all redshifts; whereas, for higher interaction strengths ($\xi, \Gamma/H_0 \,{\gtrsim}\, 0.4$) there is a possibility (in principle) for detecting the given signal at $z_S \,{\gtrsim}\, 0.5$ (for $\ell \,{\gtrsim}\, 20$). The plots also reveal the separation between $Q^{(1)}_x$ and $Q^{(2)}_x$, which increases with increasing source redshift. As previously pointed out, our normalization of the IDE models will cause them to converge as source redshift decreases, approaching the present epoch: this is revealed in the plots.


\subsection{The imprint of IDE}
\label{subsec:IDECls}
Here we discuss the effect of IDE in the RSDs angular power spectrum $\hat{C}_\ell$ (being $C_\ell$ with $\Delta_{\rm loc} \,{=}\, 0$). We look at the changes induced in $\hat{C}_\ell$ by the behaviour of the IDE, as specified by $Q^{(1)}_x$ and $Q^{(2)}_x$, respectively, by taking the ratios of $\hat{C}_\ell$ with IDE ($\Gamma/H_0 \,{\neq}\, 0$ and $\xi \,{\neq}\, 0$) to that without IDE ($\Gamma/H_0 \,{=}\, \xi \,{=}\, 0$).

In Fig.~\ref{fig:clsWRTlTotalIDEeffect} (bottom panels), we show the ratios of the RSDs angular power spectrum $\hat{C}_\ell$ (that with IDE to that without IDE), i.e. $\hat{C}_\ell^{(\Gamma/H_0 \,{\neq}\, 0,\, \xi \,{\neq}\, 0)} / \hat{C}_\ell^{(\Gamma/H_0 \,{=}\, \xi \,{=}\, 0)}$, as a function of multipole $\ell$ at source redshifts $z_S \,{=}\, 0.1,\, 1$ (accordingly), correspoonding to the plots in Fig.~\ref{fig:clsWRTlTotalIDEeffect} (top panels). We see that at $z_S \,{=}\, 0.1$, $Q^{(1)}_x$ gives relatively larger suppression at $\ell \,{\lesssim}\, 20$; whereas, at $z_S \,{=}\, 1$, $Q^{(2)}_x$ gives the larger suppression at all $\ell$. We also see that $Q^{(1)}_x$ leads to constant ratios of the RSDs angular power spectrum at $z_S \,{=}\, 0.1$, while at $z_S \,{=}\, 1$ it gives a scale-dependent behaviour. On the other hand, $Q^{(2)}_x$ always leads to a scale-dependent behaviour. Thus, it implies that $Q^{(1)}_x$ possesses an alternating (negative-positive) effect in RSDs, with respect to redshift. This could be understood by looking at the ``effective'' equation of state parameters $w_{A,\rm eff}$ \eqref{eff:EoS}, which encode the deviations from standard evolution in the dark sector energy densities. In particular, we look at the parameter $q^{(I)}_x \,{\equiv}\, a\bar{Q}^{(I)}_x/(3{\cal H}\bar{\rho}_x)$ in $w_{x,\rm eff}$: for $Q^{(2)}_x$ we have $q^{(2)}_x \,{=}\, \xi/3$, which is an absolute constant, and for $Q^{(1)}_x$ we have $q^{(1)}_x \,{=}\, a\Gamma/(3{\cal H})$, which is time-dependent. Thus, the deviation from standard evolution of DE $Q^{(2)}_x$ will remain the same at all redshifts; whereas, that of $Q^{(1)}_x$ can change at different redshifts and hence giving different imprints in RSDs at low and at high source redshifts, respectively. 

Moreover, at both source redshifts, we notice a strong sensitivity in the ratios to changes in the values of the interaction strength, for a given IDE model. We also see that for a given IDE model, the amount of separation between the ratios for the values of the interaction strength is substantial. This sensitivity and separation suggest that, in view of multi-tracer analysis, RSDs hold the potential to detect the imprint of IDE on large scales, at source redshifts $z_S \,{\lesssim}\, 1$. Furthermore, although the amplitude of the ratios at $z_S \,{=}\, 0.1$ are larger than those at $z_S \,{=}\, 1$, the separation between the ratios for $Q^{(1)}_x$ and  $Q^{(2)}_x$ at $z_S \,{=}\, 1$ is larger than those at $z_S \,{=}\, 0.1$. This suggests that, given our normalization, RSDs hold the potential to distinguish IDE models suitably at $z_S \,{\simeq} 1$, on all scales: in the light of multi-tracer analysis.

Fig.~\ref{fig:clsWRTzTotalIDEeffect} (right panel) gives the corresponding ratios of the RSDs angular power spectrum $\hat{C}_\ell$ (that with IDE to that without IDE), i.e. $\hat{C}_\ell^{(\Gamma/H_0 \,{\neq}\, 0,\, \xi \,{\neq}\, 0)} / \hat{C}_\ell^{(\Gamma/H_0 \,{=}\, \xi \,{=}\, 0)}$, as a function of source redshift $z_S$ ($0.1 \leq z_S \leq 1$) on the scale $\ell \,{=}\, 20$. These ratios correspoond to the plots in Fig.~\ref{fig:clsWRTzTotalIDEeffect} (left panel). The ratios of $\hat{C}_\ell(z_S)$ show that $Q^{(2)}_x$ leads to larger suppressions than $Q^{(1)}_x$ within $0.1 \,{\leq}\, z_S \,{\leq}\, 1$, on the given scale ($\ell \,{=}\, 20$). The separation between the ratios increases as source redshift increases, and becomes substantial at $z_S \,{\gtrsim}\, 0.5$. Thus, by taking multi-tracer analysis into account, RSDs hold the potential of distinguishing IDE models, albeit at higher source redshifts ($z_S \,{\gtrsim}\, 0.5$), on the given scale ($\ell \,{=}\, 20$).


\subsection{The cross-bin angular correlations}
\label{subsec:crossCorrCls}
In Figs.~\ref{fig:clsWRTlTotalIDEeffect} and~\ref{fig:clsWRTzTotalIDEeffect}, we looked at the RSDs auto-correlation ($\hat{z}_S \,{=}\, z_S$) angular power spectrum with respect to multipole at fixed redshifts, and also with respect to redshift at a fixed multipole. Here we look at the redshift-bin cross-correlation angular power spectrum (cross-correlation, henceforth).

\begin{figure*}\centering
\includegraphics[scale=0.3]{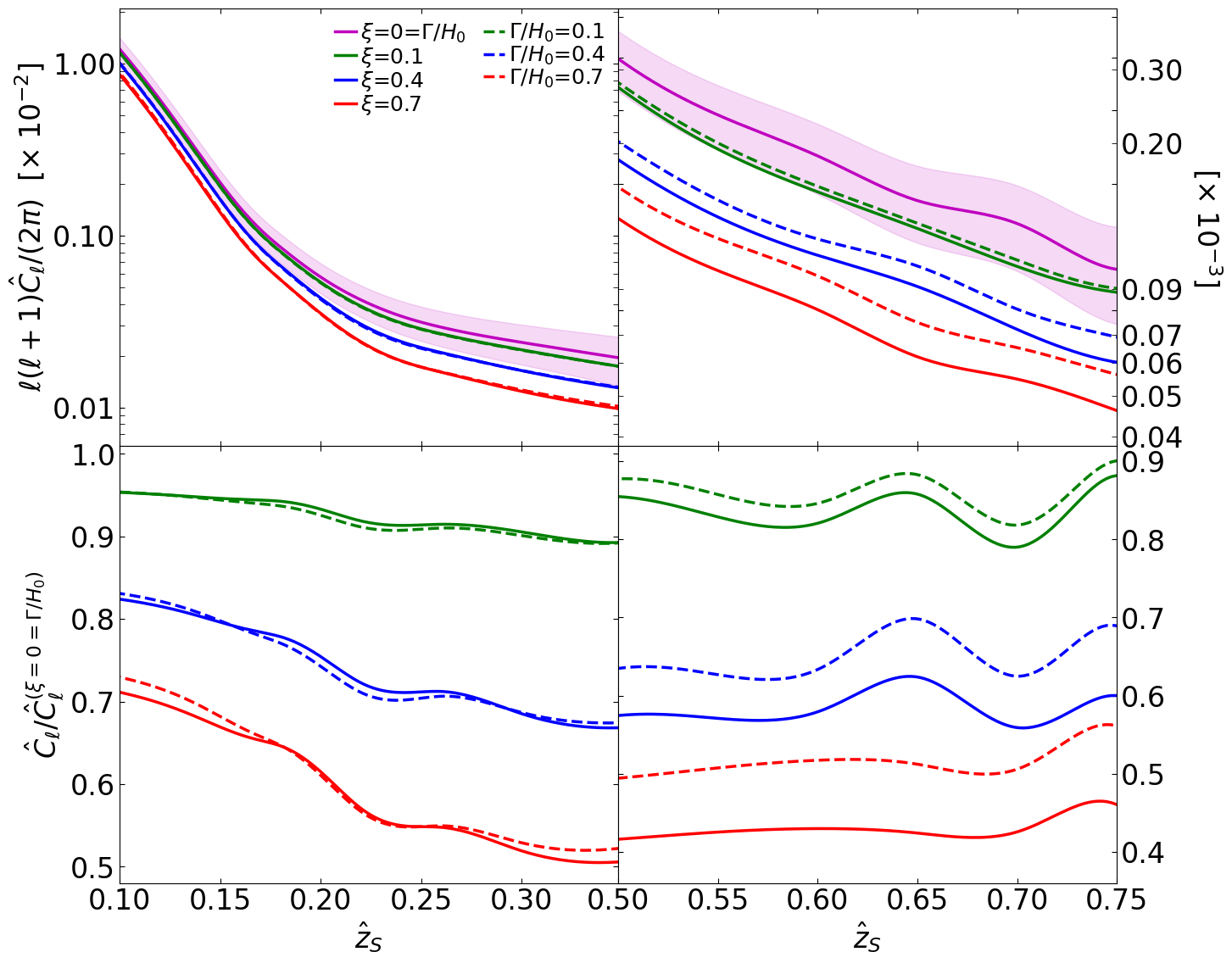}
\caption{\emph{Top:} The plots of the redshift-bin cross-correlation angular power spectrum $\hat{C}_\ell(z_S,\hat{z}_S)$ as a function of source redshift $\hat{z}_S$, for multipole $\ell \,{=}\, 20$. A bin width $\Delta{z} \,{=}\, 0.1$ and a window size $\sigma_z \,{=}\, 0.1z$ were used. The plots are at source redshifts $z_S \,{=}\, 0.1$ (left panel) and $z_S \,{=}\, 1$ (right panel). \emph{Bottom:} The plots of the corresponding ratios of $\hat{C}_\ell(z_S,\hat{z}_S)$ with IDE ($\Gamma/H_0 \,{\neq}\, 0$; $\xi \,{\neq}\, 0$) to $\hat{C}_\ell(z_S,\hat{z}_S)$ with no IDE ($\Gamma/H_0 \,{=}\, \xi \,{=}\, 0$). Note that in Figs.~\ref{fig:clsWRTlTotalIDEeffect} and~\ref{fig:clsWRTzTotalIDEeffect} we have $\hat{z}_S \,{=}\, z_S$, and $F_\ell(k,z_S) \to f_\ell(k,z_S)$.}\label{fig:clsWRTzCorrsTotalIDEeffect}
\end{figure*} 

Fig.~\ref{fig:clsWRTzCorrsTotalIDEeffect} (top panels) shows the plots of the RSDs cross-correlation $\hat{C}_\ell(z_S,\hat{z}_S)$ as a function of the second source redshift $\hat{z}_S$, on the scale $\ell \,{=}\, 20$. The extent of cosmic variance \eqref{cosmicVar} is also indicated. We computed $\hat{C}_\ell(z_S,\hat{z}_S)$ using a redshift-bin width $\Delta{z} \,{=}\, 0.1$ and standard deviation $\sigma_z \,{=}\, 0.1z$, at two initial source redshifts: $z_S \,{=}\, 0.1$ (left), with $0.1 \,{\leq}\, \hat{z}_S \,{\leq}\, 0.35$, and $z_S \,{=}\, 0.5$ (right), with $0.5 \,{\leq}\, \hat{z}_S \,{\leq}\, 0.75$. (A bin width of $\Delta{z} \,{=}\, 0.24$ has been shown~\cite{Abidi:2022zyd} to give optimal signal-to-noise ratio for spectroscopic surveys, on the scale $\ell \,{=}\, 100$.) Moreover, we made 10 redshift bins in each interval of $\hat{z}_S$. We see that the cross-correlations in the interval $0.1 \,{\leq}\, \hat{z}_S \,{\leq}\, 0.35$ (top left) look similar to the auto-correlations $\hat{C}_\ell(z_S)$ in Fig.~\ref{fig:clsWRTzTotalIDEeffect} (left panel), except that the effect of IDE for both $Q^{(1)}_x$ and $Q^{(2)}_x$ is relatively less prominent in $\hat{C}_\ell(z_S,\hat{z}_S)$. Thus, similar discussion follows. Furthermore, the amplitude of $\hat{C}_\ell(z_S,\hat{z}_S)$ appears to be lower than that of $\hat{C}_\ell(z_S)$. This can be be understood as the effect of the window function \eqref{windowfunc}. It implies that the Gaussian window function may not be suitable for analysis of phenomena in the large-scale structure, or that the adopted $\Delta{z}$ and $\sigma_z$ were not optimal. Thus, these need to be taken into account in a proper quntitative analysis. At higher redshift interval, $0.5 \,{\leq}\, \hat{z}_S \,{\leq}\, 0.75$ (top right), the cross-correlations look completely different; with apparent oscillations surfacing. These oscillations do not appear in $\hat{C}_\ell(z_S)$ (see Fig.~\ref{fig:clsWRTzTotalIDEeffect}, left panel). The oscillations may not be physical, but rather the domination of the relatively weak RSDs signal in $\hat{C}_\ell(z_S,\hat{z}_S)$ by the Bessel sperical function, given the afore mentioned (redshift) window effect. (A rigorous quantitative analysis may be required to confirm this.) 

Nevertheless, we see that at the interval $0.5 \,{\leq}\, \hat{z}_S \,{\leq}\, 0.75$ most of the plots of $\hat{C}_\ell(z_S,\hat{z}_S)$, i.e. for $(\xi,\, \Gamma/H_0) \,{\gtrsim}\, 0.1$, fall outside the cosmic variance (albeit marginally for $\xi \,{=}\, \Gamma/H_0 \,{\sim}\, 0.1$) as opposed to the corresponding plots of $\hat{C}_\ell(z_S)$ at the same redshift interval (see Fig.~\ref{fig:clsWRTzTotalIDEeffect}, left panel). This implies that the cross-bin correlations will naturally, without multi-tracer analysis, (in principle) alleviate the severity of cosmic variance---relative to the auto-correlation. Thus, taking multi-tracer analysis and other quantitative measures properly into account, the RSDs cross-correlations hold the potential of detecting the imprint of IDE, at the given $\hat{z}_S$ (for the given $\ell$). 

Moreover, Fig.~\ref{fig:clsWRTzCorrsTotalIDEeffect} (bottom panels) shows the corresponding ratios of $\hat{C}_\ell(z_S,\hat{z}_S)$ with IDE to $\hat{C}_\ell(z_S,\hat{z}_S)$ with no IDE, for the same $\ell$ and $z_S$ intervals. We see that in the lower $z_S$ interval, $0.1 \,{\leq}\, \hat{z}_S \,{\leq}\, 0.35$ (bottom left), the effects of $Q^{(1)}_x$ and $Q^{(2)}_x$ are barely differentiable. This is contrary to what is seen for $\hat{C}_\ell(z_S)$
in Fig.~\ref{fig:clsWRTzTotalIDEeffect} (right panel), where the two IDE models are clearly seen to gradually diverge at $z_S \,{>}\, 0.1$. Moreover, their are weak oscillations appearing in the $\hat{C}_\ell(z_S,\hat{z}_S)$ ratios. The behaviour of the given ratios of $\hat{C}_\ell(z_S,\hat{z}_S)$ can be attributed to the same window effect in $\hat{C}_\ell(z_S,\hat{z}_S)$ (top left). The amplitudes of the effects of the IDE models, and their strength to diverge or maintain a monotonic pattern (as seen for the same $z_S$ interval in Fig.~\ref{fig:clsWRTzTotalIDEeffect}), have been ``washed'' out by the window specifications. Thus, the IDE imprint can be lost to window effects in the large-scale cross-correlations if proper care is not taken to incorporate the right quantitative specifications. This could eventually lead to the inability to identify the imprint of IDE in constraints.

Unlike at $z_S \,{=}\, 0.1$ ($0.1 \,{\leq}\, \hat{z}_S \,{\leq}\, 0.35$), where oscillations only surface in the $\hat{C}_\ell(z_S,\hat{z}_S)$ ratios, we see for $z_S \,{=}\, 0.5$ ($0.5 \,{\leq}\, \hat{z}_S \,{\leq}\, 0.75$) these oscillations appear in both $\hat{C}_\ell(z_S,\hat{z}_S)$ and the ratios (in which the oscillations become more apparent at all $\hat{z}_S$), and also in the separation between the IDE models. The difference between the ratios for the two IDE models appears to be relatively small for $(\xi,\, \Gamma/H_0) \,{\lesssim}\, 0.1$; however, it gradually becomes significant for higher values of $\xi$, and $\Gamma/H_0$. On the other hand, for each IDE model, the separation between the lines for successive values of the interaction strength, is substantial at all $\hat{z}_S$ (for dashed and for solid lines, respectively). This implies that $\hat{C}_\ell(z_S,\hat{z}_S)$ is sensitive to subtle changes in the IDE interaction strength. Thus, given multi-tracer analysis, the RSDs cross-correlations hold the potential of probing the nature of IDE at all $\hat{z}_S$ in the given interval.


\subsection{The RSDs signal in IDE}
\label{subsec:RSDSignal}

In Secs.~\ref{subsec:Cls}--\ref{subsec:crossCorrCls}, we only considered the density-RSDs angular power spectrum $\hat{C}_\ell \,{=}\,  C^{(\Delta_{\rm loc} {=}\, 0)}_\ell$, i.e. the angular power spectrum with the local term \eqref{Delta-loc} being neglected. However, as previously pointed out (Sec.~\ref{subsec:Dens}), in reality RSDs exist among several effects other than the matter density; these include local effects (considering only non-integral effects). Although the local effects will typically amount to a relatively small contribution compared to the terms in \eqref{Delta-RSD}, they need to be taken into account in the light of upcoming precision cosmology.

However, when the local effects are included, the resulting angular power spectrum can no longer be taken as the RSDs angular power spectrum but the total galaxy angular power spectrum of low-redshift regime. Nevertheless, we are able to estimate the RSDs signal from the total (galaxy) angular power spectrum, given by
\beq\label{RSD-effect}
(\rm RSD\, Signal)_\ell \;\equiv\; \dfrac{C_\ell - C^{(\rm no\, RSD)}_\ell}{C^{(\rm no\, RSD)}_\ell},
\eeq
where $C_\ell$ is the full angular power spectrum \eqref{Cls} prescribed by \eqref{f_ell}, and $C^{(\rm no\, RSD)}_\ell$ is the angular power spectrum with the second term (first line) in \eqref{f_ell} being neglected (or set to zero). The main goal is to understand the behaviour of the RSDs signal; thus, only the auto-correlation ($\hat{z}_S \,{=}\, z_S$) total angular power spectrum $C_\ell$, as a function of $\ell$ (at fixed $z_S$) and as a function of $z_S$ (at fixed $\ell$), is considered.

\begin{figure*}\centering
\includegraphics[scale=0.32]{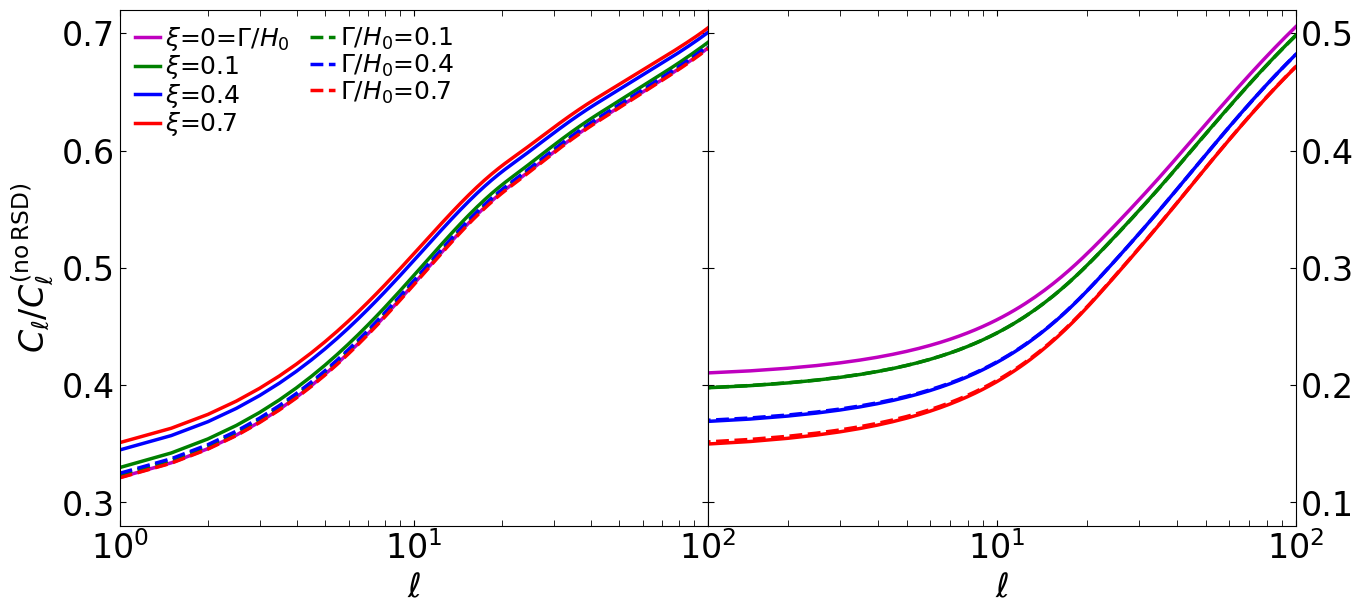}
\caption{The plots of the ratio $C_\ell/C^{(\rm no RSD)}_\ell$ as a function of multipole $\ell$, where $C_\ell$ is the total angular power spectrum given by \eqref{Cls} and \eqref{f_ell}, and $C^{(\rm no\, RSD)}_\ell$ is the angular power spectrum with the second term in the first line in \eqref{f_ell} being neglected. The panels are for $z_S \,{=}\, 0.1$ (left) and $z_S \,{=}\, 1$ (right). All notations are as in Fig.~\ref{fig:clsWRTlTotalIDEeffect}. These ratios measure $1+(\rm RSD\, Signal)_\ell$.}\label{fig:clsWRTlRSDeffect}
\end{figure*}

Fig.~\ref{fig:clsWRTlRSDeffect} shows plots of $C_\ell/C^{(\rm no RSD)}_\ell \,{=}\, 1+(\rm RSD\, Signal)_\ell$ as a function of $\ell$, at $z_S \,{=}\, 0.1$ (left panel) and $z_S \,{=}\, 1$ (right panel). The RSDs signal estimated this way \eqref{RSD-effect} properly incorporates the full contribution of the other terms (matter density and local effects) in the angular power spectrum. We see that the amplitude of the ratios decreases steeply with increasing scale (decreasing $\ell$) at both $z_S$, with a gradual tendency toward flattening at $\ell \,{\lesssim}\, 10$. Moreover, the amplitude of the ratios is higher at $z_S \,{=}\, 0.1$ and at $z_S \,{=}\, 1$. Thus, it implies that the RSDs signal \eqref{RSD-effect} diminishes quickly on the largest scales, with its amplitude slowly decreasing with increasing redshifts. Furthermore, the ratios for $Q^{(1)}_x$ and $Q^{(2)}_x$ appear to coincide at both $z_S$, with the separation between successive ratios (values of interaction strength) becoming relatively prominent at $z_S \,{=}\, 1$. It should be noted that the increase in separation between successive ratios, for a given IDE model, at $z_S \,{=}\, 1$ has nothing to do with our normalization: the normalization only determines the separation between the IDE models themselves (i.e. solid lines; dashed lines), for a given value of $\xi \,{=}\, \Gamma/H_0$. However, the increase in separation between successive ratios at $z_S \,{=}\, 1$ suggests that RSDs become more sensitive to changes in IDE at higher redshifts, and hence will be suitable for probing the nature and imprint of IDE at the given $z_S$.

\begin{figure*}\centering
\includegraphics[scale=0.32]{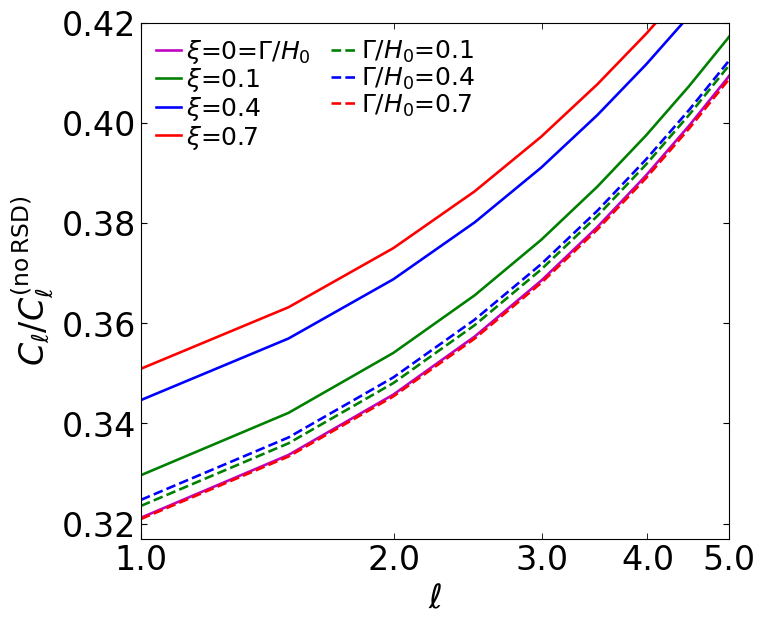} \includegraphics[scale=0.32]{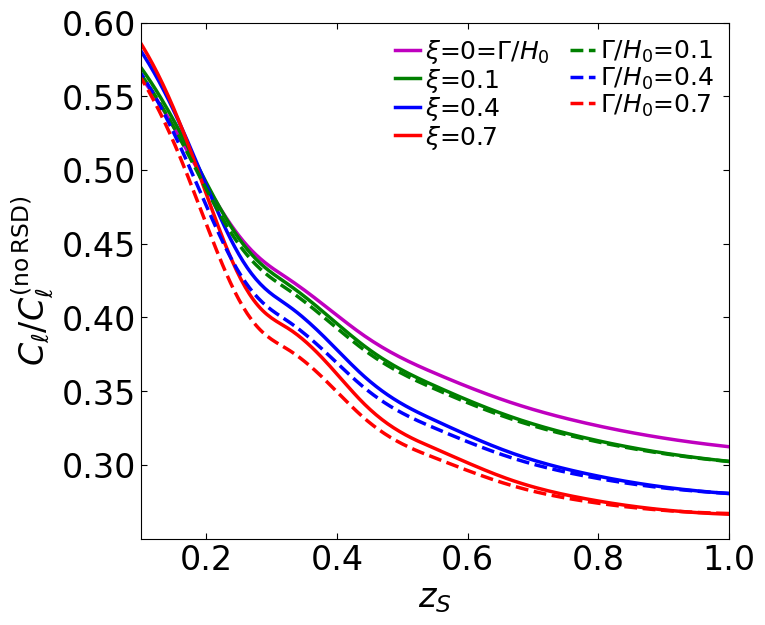}
\caption{\emph{Left:} A zoom-in of the plots in the left panel of Fig.~\ref{fig:clsWRTlRSDeffect}. \emph{Right:} The plots of the ratio $C_\ell/C^{(\rm no RSD)}_\ell$ as a function of source redshift $z_S$, on the scale $\ell \,{=}\, 20$. All notations are as in Figs.~\ref{fig:clsWRTlTotalIDEeffect} and~\ref{fig:clsWRTlRSDeffect}. As in Fig.~\ref{fig:clsWRTlRSDeffect}, these ratios measure $1+(\rm RSD\, Signal)_\ell$, with $(\rm RSD\, Signal)_\ell$ being given by \eqref{RSD-effect}.}\label{fig:clsWRTzRSDeffect}
\end{figure*}

We also observe that at $z_S \,{=}\, 0.1$ (left panel), $Q^{(2)}_x$ leads to a consistent amplitude enhancement of the ratios with increasing values of $\xi$; whereas, $Q^{(1)}_x$ leads to a somewhat suppression or oscillatory behaviour---a slight increase for increasing values of $\Gamma/H_0 \,{\leq}\, 0.4$, and then a decrease for $\Gamma/H_0 \,{>}\, 0.4$. (A zoomed-in image of these results is given in Fig.~\ref{fig:clsWRTzRSDeffect}, left panel, for appreciation.) This is also consistent with findings in Sec.~\ref{subsec:IDECls}, where $Q^{(1)}_x$ leads to a positive-negative alternating effect, at the same $z_S$. At $z_S \,{=}\, 1$, both IDE models give a consisten amplitude suppression with increasing interaction strengths. Moreover, we notice that at both source redshifts (Fig.~\ref{fig:clsWRTlRSDeffect}), we have $(\rm RSD\, Signal)_\ell \,{<}\, 0$, which implies that $C_\ell \,{<}\, C^{(\rm no RSD)}_\ell$. This shows that RSDs will combine with the density amplitude and local effects in the observed overdensity to give a negative contribution in the total angular power spectrum, at the given $z_S$ and $\ell$.

Fig.~\ref{fig:clsWRTzRSDeffect} (right panel) shows the plots of the same ratios in Fig.~\ref{fig:clsWRTlRSDeffect}, but here as functions of source redshift $z_S$. We see that both IDE models give consistent amplitude separation with increasing interaction strengths. The models appear to coincide on majority of the redshift interval, except for $ 0.2 \,{\lesssim}\, z_S \,{\lesssim}\, 0.6$ and $\xi \,{=}\, \Gamma/H_0 \,{\gtrsim}\, 0.4$ where the models appear to separate slightly. In general, we see that the RSDs signal diminishes as $z_S$ increases. Moreover, as in Fig.~\ref{fig:clsWRTlRSDeffect}, we have $(\rm RSD\, Signal)_\ell \,{<}\, 0$, implying that RSDs will combine with other effects and give a negative contribution in the total angular power spectrum on the given $z_S$ interval, for the given $\ell$.

Note that the actual plots of $C_\ell$ are almost identical with those of $\hat{C}_\ell$, hence the plots of $C_\ell$ are not shown. However, for completeness, the plots of the fractional change between the two are given as a function of $\ell$ in Fig.~\ref{fig:clsWRTlLocEffect}. The plots show a (positive) contribution by the local effects of less than $6\%$ at $z_S \,{=}\, 0.1$, while at $z_S \,{=}\, 1$ they are at sub-percent level. (See also discussion on Fig.~\ref{fig:clsWRTlLocEffect} in Appendix~\ref{app:LocSignal}.) As previously stated, although some comparisons are made between the given IDE models, the goal of this work is not to compare them but to present the analysis in more than one model (for appreciation); hence using two most widely investigated IDE models.


\subsection{Imprint of clustering and evolution biases}
\label{subsec:galEvolbias}

Bias occurs in the clustering of galaxies, and also in their evolution. The two biases should ideally be related (see e.g.~Refs.~\refcite{Duniya:2016ibg} and~\refcite{Jeong:2011as} for analytical relations between the two biases). The former, arises as a result of the imperfect mapping of the underlying matter density by the galaxy number-density distribution in a given region; whereas, the latter arises from the inability of the galaxy number-density to accurately track and trace the cosmic progression of the underlying matter density over time. (See Ref.~\refcite{Desjacques:2016bnm} for an extensive review on clustering bias.) Here we try to gain some insights into the effect of biasing in RSDs.

\begin{figure*}\centering
\includegraphics[scale=0.26]{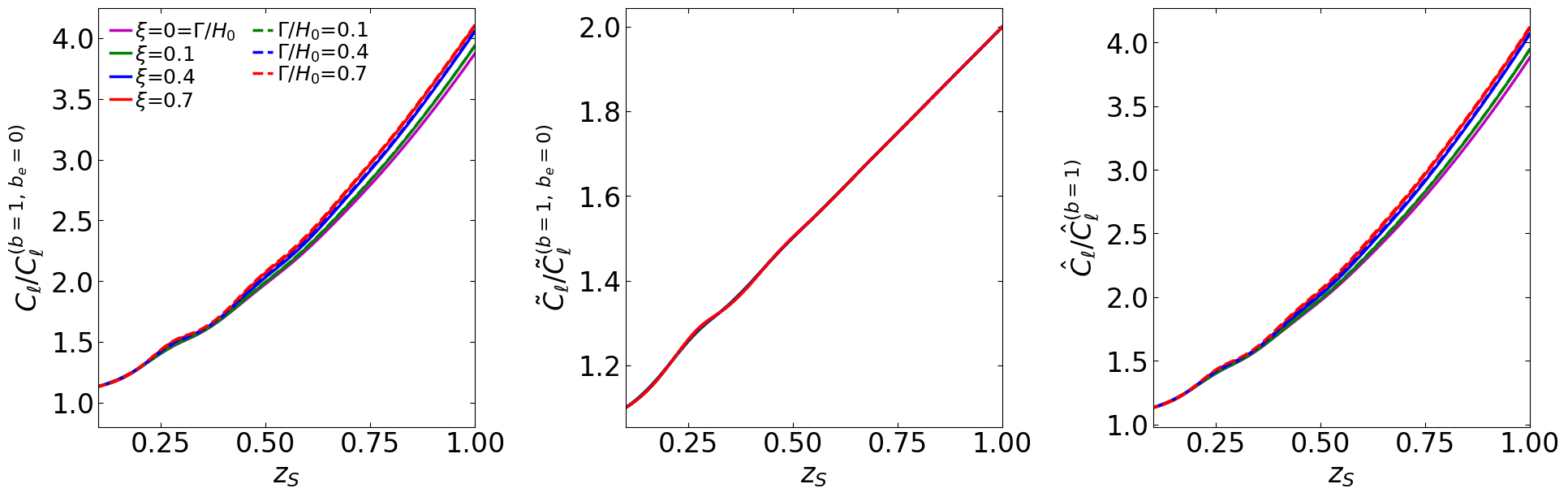}
\caption{The plots of the ratios of the total angular power spectrum $C_\ell$, the local-effect angular power spectrum $\tilde{C}_\ell \,{=}\, C^{(\rm no\, RSD)}_\ell$, and the RSDs angular power spectrum $\hat{C}_\ell$ (see Figs.~\ref{fig:clsWRTlTotalIDEeffect}--\ref{fig:clsWRTzCorrsTotalIDEeffect}), as functions of source redshift $z_S$. The angular power spectra $C_\ell$, $\tilde{C}_\ell$ and $\hat{C}_\ell$ have $b$ given by \eqref{b-fit} and, $C_\ell$, $\tilde{C}_\ell$ have $b_e$ given by \eqref{b_e2} and \eqref{N-fit} ($\hat{C}_\ell$ has no $b_e$, since it has $\Delta_{\rm loc} \,{=}\, 0$). The denomenators in the $y$-axes indicate the parameter specifications applied. Other notations are as given in previous figures.}\label{fig:clsWRTzbbeEffects}
\end{figure*}

Fig.~\ref{fig:clsWRTzbbeEffects} shows the plots of the ratios of the total angular power spectrum $C_\ell$, the ``local-effect'' angular power spectrum $\tilde{C}_\ell \,{=}\, C^{(\rm no\, RSD)}_\ell$, and the RSDs angular power spectrum $\hat{C}_\ell$ by their modified versions $C^{(b{=}1,\, b_e{=}0)}_\ell$, $\tilde{C}^{(b{=}1,\, b_e{=}0)}_\ell$ and $\tilde{C}^{(b{=}1)}_\ell$, respectively, as functions of source redshift $z_S$ on the scale $\ell \,{=}\, 20$: $C_\ell$ and $\tilde{C}_\ell$ were computed using the clustering bias \eqref{b-fit} and the evolution bias, given by \eqref{b_e2} and \eqref{N-fit}, and $\hat{C}_\ell$ was computed with only the clustering bias \eqref{b-fit} (since $\Delta_{\rm loc} \,{=}\, 0$ in $\hat{C}_\ell$). The plots measure the effect of biasing (clustering and evolutionary) in the large-scale structure, under the influence of different IDE strengths, for a given IDE type. We see that, in general, neither of the IDE models results in any appreciable impact. Nevertheless, the plots reveal a few insights: Firstly, the biases lead to a positive net effect in the angular power spectra, with $(C_\ell - C^{(b{=}1,\, b_e{=}0)}_\ell) / C^{(b{=}1,\, b_e{=}0)}_\ell \,{>}\, 0$ (same for $\tilde{C}_\ell$ and $\hat{C}_\ell$). Secondly, the presence of RSDs tends to endow the angular power spectrum with the potential (albeit marginal) to detect the imprint (or presence) of IDE, as the absence of RSDs (middle panel) results in the erasure of the IDE effect for all interaction strengths---and the biasing signal for IDE all converge on that of (standard) non-IDE. However, including RSDs (left and right panels, respectively) causes the biasing signal for IDE to show signs of gradual deviations from that for non-IDE, as $z_S$ increases. Thus, proper modelling of bias can play a crucial role in the analysis of IDE in the large-scale structure.

In Secs.~\ref{subsec:Cls}--\ref{subsec:galEvolbias}, we considered the galaxy (clustering and evolution) biases to be independent of the dark-sector interactions; hence used \eqref{b-fit} and \eqref{N-fit} of standard, non-IDE analysis. In what follows, we consider a scenario where the two biases are incorporate the IDE dynamics.


\subsection{Dark-sector interaction in galaxy biases}
\label{subsec:galevolbias-IDE}

\begin{figure*}\centering
\includegraphics[scale=0.3]{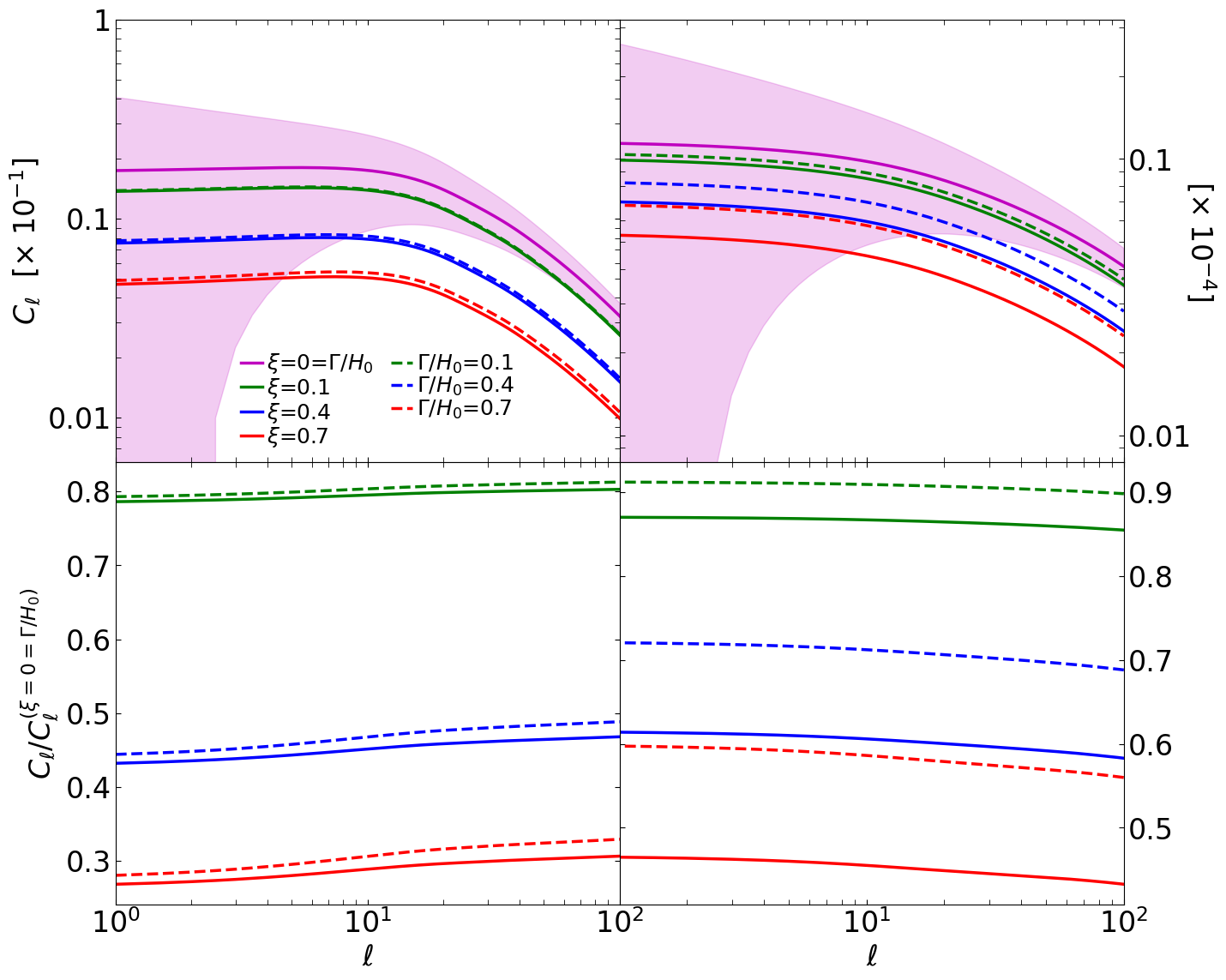}
\caption{\emph{Top:} The plots of the total angular power spectrum $C_\ell$ as a function of multipole $\ell$, at fixed source redshifts $z_S \,{=}\, 0.1$ (left) and $z_S \,{=}\, 1$ (right): the clustering and the evolution biases are as given in Sec.~\ref{subsec:galevolbias-IDE}, where $b_e \,{\to}\, \check{b}_e(\rm IDE)$ and $b \,{\to}\, \check{b} \,{=}\, \check{b}(\check{b}_e,w_{m,\rm eff})$. The overhead check symbol denotes dependence on IDE dynamics. For simplicity, we set $\check{b}_e(z) \,{=}\, b_e(z)$, where $b_e$ is as given by \eqref{b_e2} and \eqref{N-fit}. \emph{Bottom:} The plots of the corresponding ratios of $C_\ell$, with IDE ($\Gamma/H_0 \,{\neq}\, 0$ and $\xi \,{\neq}\, 0$) to that with no IDE ($\Gamma/H_0 \,{=}\, 0 \,{=}\, \xi$), at the given $z_S$. All notations are as in Fig.~\ref{fig:clsWRTlTotalIDEeffect}.}\label{fig:clsWRTlTotalbFrombeIDEeffect}
\end{figure*}

When the dark-sector components interact in the background universe, i.e. there is an exchange of energy (momentum exchange only affects the perturbations), then the redshift-dependent galaxy (clustering and evolution) biases should naturally acquire the background dark-sector interaction via the evolving energy densities and the associated parameters. (Momentum transfer will be important for bias models with scale dependence or up to at least first-order in perturbations.) This interaction needs to be taken into account in modelling the biases. This has never previously been considered; with the fitting formulars for non-IDE being taken as standard. However, this is inconsistent, and could lead to incorrect estimation of IDE imprint in constraints.

Here we make a first attempt at including the IDE dynamics in the biases. Refs.~\refcite{Duniya:2016ibg} and~\refcite{Jeong:2011as} present simple expressions relating the galaxy biases, respectively, for the non-IDE scenario in the form, $b \,{=}\, \alpha_0 + \alpha_1 b_e$, where $\alpha_0$ and $\alpha_1$ are constants. However, in general (including IDE and modified gravity), the biases take different forms ($b \,{\to}\, \check{b}$ and $b_e \,{\to}\, \check{b}_e$), and are related by \cite{Duniya:2016ibg}
\beq\label{galbias-IDE}
\check{b} \,=\, -{\cal H}\left(3 - \check{b}_e\right) \dfrac{\bar{\rho}_m}{\bar{\rho}'_m} \;=\; \dfrac{3 - \check{b}_e}{3\left(1+w_{m,\rm eff}\right)},
\eeq
where $\bar{\rho}_m$ and $\bar{\rho}'_m$ are as given by \eqref{conservatn}, and $w_{m,\rm eff}$ is as given by \eqref{eff:EoS}. (Note that for non-IDE, $w_{m,\rm eff} \,{=}\, w_m \,{=}\, 0$ and, we have $\alpha_0 \,{=}\, 1$ and $\alpha_1 \,{=}\, {-}1/3$.) To demonstrate the effect of incorporating IDE dynamics in the biases, we only consider the total angular power spectrum as a function of multipole, at given source redshifts. For simplicity, we take $\check{b}_e(z) \,{=}\, b_e(z)$ with $b_e$ being as prescribed by \eqref{b_e2} and \eqref{N-fit}, and we use \eqref{galbias-IDE} for the clustering bias. Note that the bias relation \eqref{galbias-IDE} can be used to obtain either of the biases once the other is known.

Fig.~\ref{fig:clsWRTlTotalbFrombeIDEeffect} (top panels) show the plots of the total angular power spectrum $C_\ell$ as a function of $\ell$, at $z_S \,{=}\, 0.1$ (left) and $z_S \,{=}\, 1$ (right); with the evolution bias, \eqref{b_e2} and \eqref{N-fit}, and the IDE-dependent clustering bias \eqref{galbias-IDE} being used. The extent of cosmic variance \eqref{cosmicVar} in $C_\ell$ is also indicated (shaded regions) in the plots. It should be pointed out that the plots of $\hat{C}_\ell$ corresponding to $C_\ell$ (Fig.~\ref{fig:clsWRTlTotalbFrombeIDEeffect}) are almost identical in form and amplitude; with the slight difference arising in amplitude on scales $\ell \,{\lesssim}\, 10$. This is similar to the scenario in Sec.~\ref{subsec:Cls} (see also Appendix~\ref{app:LocSignal}); hence the plots of $\hat{C}_\ell$ with respect to \eqref{galbias-IDE} are not shown. 

As expected, we observe in Fig.~\ref{fig:clsWRTlTotalbFrombeIDEeffect} (top panels) the usual suppression of power in $\hat{C}_\ell$ as interaction strength increases, for both $Q^{(1)}_x$ (dashed lines) and $Q^{(2)}_x$ (solid lines), at both the given source redshufts. Moreover, we see the plots for $Q^{(1)}_x$ appear to almost overlap with those of $Q^{(2)}_x$ at $z_S \,{=}\, 0.1$, but become noticeably separated at $z_S \,{=}\, 1$; with the separation increasing with increase in the values of $\xi$ and $\Gamma/H_0$. One reason for this will be our normalization. (The IDE models will converge as $z_S \,{\to}\, 0$, and diverge otherwise.) On the other hand, the results show that the separation between the plots for each value of the interaction strength, for a given IDE model, is relatively larger at $z_S \,{=}\, 0.1$ and at $z_S \,{=}\, 1$. From the given results, it suggests that the RSDs angular power spectrum is more sensitive to changes in the behaviour of IDE at $z_S \,{=}\, 0.1$ than at $z_S \,{=}\, 1$, and hence will be better for analysing the imprint of IDE and placing constraints on a given IDE model; whereas high redshifts will be suitable for distinguishing IDE models (though this may be only subject to our normalization).

Fig.~\ref{fig:clsWRTlTotalbFrombeIDEeffect} (bottom panels) show the plots of the corresponding ratios of $C_\ell$---with IDE ($\Gamma/H_0 \,{\neq}\, 0$ and $\xi \,{\neq}\, 0$) to that with no IDE ($\Gamma/H_0 \,{=}\, 0 \,{=}\, \xi$)---as a function of $\ell$, at the given $z_S$. As previously indicated by the plots of $C_\ell$ (top panels), the ratios for $Q^{(1)}_x$ and $Q^{(2)}_x$ are of the same order of magnitude for all the values of $(\xi,\, \Gamma/H_0) \,{>}\, 0$ at $z_S \,{=}\, 0.1$, while at $z_S \,{=}\, 1$ they are strongly diverging with increasing magnitude of the values of $(\xi,\, \Gamma/H_0) \,{>}\, 0$. Similarly, for a given IDE model, although the separation between the ratios is larger at $z_S \,{=}\, 0.1$ than at $z_S \,{=}\, 1$, it is nevertheless substantial at both source redshifts; also with the separations at $z_S \,{=}\, 0.1$ being larger here than in Fig.~\ref{fig:clsWRTlTotalIDEeffect} (bottom left), where the IDE dynamics is not included in the biases. The IDE imprint is more prominent when its dynamics is incorporated in the biases. Moreover, the apparent positive-negative alternating behaviour in $Q^{(1)}_x$ which is seen in Fig.~\ref{fig:clsWRTlTotalIDEeffect} completely disappears in Fig.~\ref{fig:clsWRTlTotalbFrombeIDEeffect}, with the given IDE model appearing to give consistent effect in the total angular power spectrum. Thus, the results in Fig.~\ref{fig:clsWRTlTotalbFrombeIDEeffect} implies that low redshifts ($z_S \,{\lesssim}\, 1$) will be suitable for constraining IDE models in RSDs and hence, identifying the imprint of IDE. However, including IDE dynamics in the galaxy biases holds better potential at constraining the imprint of IDE, since sensitivity to changes in the IDE parameters is higher. Also, although the results reveal a significant deviation between the given IDE models at $z_S \,{=}\, 1$, it is not conclusive whether the given epoch will be suitable for distinguishing IDE models: this may be solely a consequence of our normalization; further investigation will be required. Furthermore, these results suggest that ignoring IDE dynamics in the galaxy biases can lead to ``artefacts'' in the analysis, and hence incorrect estimation of the imprint of IDE in the cosmological parameters.


\section{Conclusion}\label{sec:Concl}
A qualitative analysis of the redshift space distortions angular power spectrum (on ultra-large scales), for two interacting dark energy models, was presented. The analysis was performed at fixed redshifts, over a redshift interval, and for (redshift $z$) cross-bin correlations. Two main scenarios were considered: The first, in which interacting dark energy dynamics were not taken into account in the galaxy (clustering and evolution) biases---as conventionally done. The second, in which interacting dark energy dynamics were incorporated in the galaxy biases.

Without taking interacting dark energy dynamics into account in the galaxy biases, the results suggested that for a constant dark energy equation of state parameter, an interacting dark energy model where the dark energy transfer rate is proportional to the dark energy density will give an alternating, positive-negaitive effect in the redshift space distortions angular power spectrum, as redshift increases. 

However, by incorporating interacting dark energy dynamics in the galaxy biases, it was found that the apparent positive-negative alternating behaviour by an interacting dark energy with a constant dark energy equation of state parameter---where the dark energy transfer rate is proportional to the dark energy density---completely disappears, with the given interacting dark energy giving a consistent effect in the angular power spectrum. This implies that ignoring interacting dark energy dynamics in the galaxy biases can lead to ``artefacts'' (unphysical signatures) in the relevant analysis; consequently, leading to incorrect identification of the imprint of interacting dark energy. Thus, proper and comprehensive modelling of the galaxy biases can enhance the true potential of redshift space distortions as a probe of interacting dark energy, and play a crucial role in the analysis of interacting dark energy in the large-scale structure.

In both scenarios, the results showed that multi-tracer analysis will be needed to beat down cosmic variance in order for the angular power spectrum as a statistic to be a viable diagnostic of redshift space distortions and interacting dark energy. Moreover, the results implied that in view of multi-tracer analysis, redshift space distortions hold the potential to constrain the imprint of interacting dark energy on very large scales, at low redshifts ($z \,{\leq}\, 1$); with the scenario having IDE dynamics incorporated in the biases showing better potential. At the same redshifts, we also found that redshift space distortions will combine with local effects in the observed overdensity to give a negative contribution in the total (galaxy) angular power spectrum.

\section*{Acknowledgments}
Thanks to the South African Centre for High Performance Computing for making their facilities available for all the numerical computations in this work. 


\appendix

\section{Signal of Local Effects}
\label{app:LocSignal}

\begin{figure*}\centering
\includegraphics[scale=0.32]{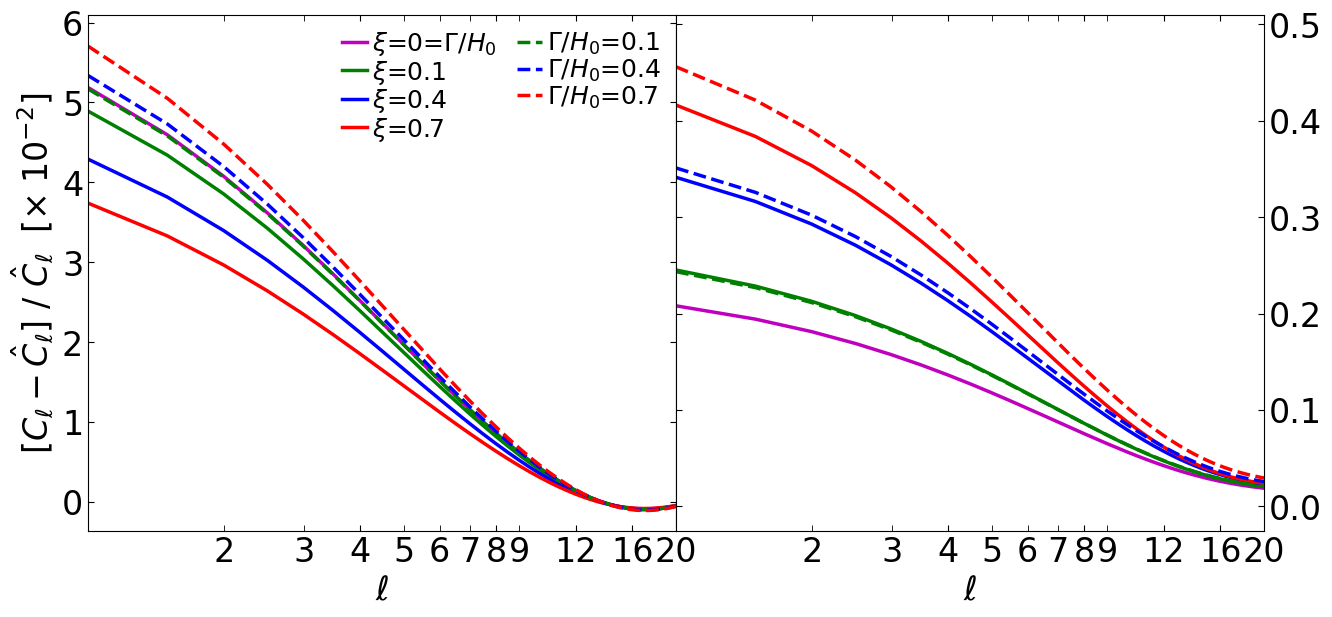}
\caption{The plots of the fractional change in the total angular power spectraum $C_\ell$ relative to the RSDs angular power spectrum $\hat{C}_\ell \,{=}\, C^{(\Delta_{\rm loc} {=} 0)}_\ell$, as a function of multipole $\ell$, at source redshifts $z_S \,{=}\, 0.1$ (left) and $z_S \,{=}\, 1$ (right). These plots are according to the specifications in Sec.~\ref{subsec:Cls}.}\label{fig:clsWRTlLocEffect}
\end{figure*}

In Fig.~\ref{fig:clsWRTlLocEffect}, we show the plots of the fractional change in the total angular power spectraum $C_\ell$ relative to the RSDs angular power spectrum $\hat{C}_\ell \,{=}\, C^{(\Delta_{\rm loc} {=} 0)}_\ell$, as a function of $\ell$, at $z_S \,{=}\, 0.1$ (left panel) and $z_S \,{=}\, 1$ (right panel). Ingeneral, the plots show a (positive) contribution by the local effects of less than $6\%$ at $z_S \,{=}\, 0.1$, while at $z_S \,{=}\, 1$ they give contributions at sub-percent level. Moreover, once again we see the positive-negative alternating effect by $Q^{(1)}_x$: at $z_S \,{=}\, 0.1$, $Q^{(1)}_x$ leads to amplitude enhancement with larger values of $\Gamma/H_0$; whereas, at $z_S \,{=}\, 1$, the model leads to amplitude suppression with larger values of $\Gamma/H_0$. On the other hand, $Q^{(2)}_x$ leads to amplitude suppression with larger values of $\xi$ at both $z_S \,{=}\, 0.1$ and $z_S \,{=}\, 1$. This supports findings in Sec.~\ref{sec:Delta-RSD}. Also, we notice that the separation between the IDE models significantly prominent at $z_S \,{=}\, 0.1$ for each value of the interaction strength, relative to that at $z_S \,{=}\, 1$. Thus, by taking the appropriate and optimal quantitative methods (which include multi-tracer analysis) into account, this suggests that the signal of local effects relative to RSDs hold the potential (in principle) to distinguish IDE models, at $z_S \,{\lesssim}\, 0.1$. The results also show that, for a given IDE model, the separation between the lines of fractional change is relatively larger at $z_S \,{=}\, 1$, which suggests that these effects become more sensitive to changes in the nature of IDE at the given $z_S$; thus, if measurable, will be important in probing the nature of IDE and placing constraints on IDE models.

\bibliographystyle{ws-ijmpd}
\bibliography{probing_IDE_with_RSD-ijmpd} 

\end{document}